Collegio Carlo Alberto

# A note on reinforcement learning to develop self-defined agents' behavior

Matteo Morini
Pietro Terna

*A note on reinforcement learning to develop self-defined agents' behavior*

Matteo Morini and Pietro Terna[¥], December 2020.

**Abstract**

The key point in this note is the self-development of simple behavior strategies, consistently with the bounded rationality hypothesis. Our artificial agents adopt learning techniques, mainly unsupervised, to achieve internal consistency in their behavior, with unexpected results. Those results can be considered mainly as the effects of the observer interpretation. The first technique in use has the name Cross Targets: to train the learning agent we use data crossed between the guesses about the action to be done and the guesses about the following results. An application of the CT "blind" strategy development is then presented: random walkers solve a node classification problem on a graph, after having learnt how to remain in homogeneous regions.

## Introduction

In the bounded rationality perspective, we have to take in deep consideration the learning phenomenon. We can investigate it via reinforcement learning techniques, mainly applied to artificial neural networks, as a way to shape agents' behavior.

A special line of work is that of inducing agents to develop their way of behaving, without using a priori (economic) rules. The section *1. Cross Targets (CT)* develops such idea and introduces the structure of an original method. The core of the technique is explained at first and then while describing the examples some sophistications are introduced, like the possibility of tuning agents' behavior.

In the section *2. Experiments* we introduce a few examples, starting with a simple analysis of the movements of an agent foraging for food. In a second example, the agent is supposed to behave in an (economic) context, via a model describing a consumer reacting to price changes.

The section *3. Advanced techniques and perspectives* introduces a significant step ahead, considering adaptive agents evolving an exploration strategy on graphs – an updated reinforcement learning technique of sorts – and introducing learning in agents to be employed in solving a classification problem inspired by standard semi-supervised learning.

[¥] Matteo Morini, Credimi S.p.A, Italia, 0000-0002-5328-8613
Pietro Terna, University of Torino, Italia, retired professor, and Fondazione Collegio Carlo Alberto, Honorary Fellow, 0000-0002-7416-0870, pietro.terna@unito.it

## 1. Cross Targets (CT)

Recent works underline the importance of the agent-based constructions to develop models of bounded rationality with learning agents (e.g., chapter 5 of Delli Gatti et al., 2018).
The critical question is: in which way we define the rules for our agents, escaping from our subjectivity?
To reply, we introduce the Cross Targets (CTs), a technique coming from previous works of one of us (starting from Terna's chapter 4 in Beltratti, Margarita and Terna, 1996). The name is related to the way we follow to figure out the learning targets in a class of models founded upon artificial adaptive agents, whose main characteristic is developing some internal consistency.
Our agents are developed via neural networks.
We can specify two types of outputs of the artificial neural network upon which we build each agent: (1) actions to be performed and (2) evaluations of the effects of those actions. With our technique, both the targets necessary to train the network from the point of view of the actions, and those connected with the effects, are built in a crossed way. The targets of type (1) are built in a consistent way with the outputs of the network concerning the guesses of the outputs of type (2), in order to develop the capability of deciding actions close to the expected results. The targets of type (2) are built in a consistent way with the outputs of the network concerning the guesses of the outputs of type (1); in this way, we will improve also the agent's capability of estimating the effects emerging from the actions that she is deciding.

### 1.1 The idea behind the mechanism

What is the idea behind such a mechanism? Our hypothesis is that an agent acting in an (economic) environment must develop and adapt her capability of evaluating in a coherent way what she has to do in order to obtain a specific result or to appreciate the consequences of a specific action. The same is true if the agent is interacting with other agents of the same population or of other populations.
The idea received some parallel conformations by an interview to Sarah-Jayne Blakemore (Zimmer, 2005):

> In recent years, scientists have moved beyond injured brains to healthy ones, thanks to advances in brain imaging. At University College London, researchers have been using brain scans to decipher how we become aware of our own bodies. «This is the very basic, low-level first point of the self» UCL's Sarah-Jayne Blakemore says.

> When our brains issue a command to move a part of our bodies, two signals are sent. One goes to the brain regions that control the particular parts of the body that need to move, and another goes to regions that monitor the movements. «I like to think of it as a 'cc' of an e-mail,» Blakemore observes. «It's all the same information sent to a different place.»

See also Blakemore and Frith (2003) about some researches in this perspective. More recently, we found idea quite close to our agent scheme, in the direction of the self-supervised prediction (Pathak, et al., 2017).

### 1.2 The learning phases

Beyond consistency, we can add other characteristics, mainly to obtain the possibility of tuning agents to make experiments.

The CTs technique attributes a central role to learning mechanisms and can be applied without introducing, either explicitly or implicitly, (economic) rules in order to influence or to characterize agents' behavior. The aim is conducting (economic) experiments without the influence of any prior (economic) hypothesis.

We will see that CTs can reproduce (economic) subjects' behavior, often in internal *ingenuous* ways, but externally with apparently complex results. The behavior, simple or complex, can appear directly as a by-product of the development of consistency between (1) decisions about actions and (2) guesses about effects. For an external observer, the agent is apparently operating with goals and plans. Obviously, it has no such symbolic entities, which are *creations* of the observer. The similarity that we recall here is that the observations and analyses about real world agents' behavior can suffer from the same bias.

The CTs algorithm introduced here is not the unique way of dealing with agents in a way consistent with the bounded rationality hypothesis, but it represents an interesting tool because it does not require injections of rules, optimizing behavior, planning capabilities, but only a limited computational ability: that one necessary to take simple decisions and to compare guesses with results.

We choose the neural approach to develop CTs mostly as a consequence of the intrinsic adaptive capabilities of neural functions. Here we will use feed forward multilayer ones.

Targets in learning process are: (1) on one side, the actual effects of the actions made by the simulated subject; (2) on the other side, the actions needed to match those guessed effects.

Figure 1 describes an agent learning and behaving in a CT scheme. The agent has to produce guesses about both its actions and their effects, on the basis of an information set (the input elements are $I_1$, ..., $I_k$). Actual effects are estimated through the guessed actions, considering also the consequences from the interaction among agents, if any; the results are used to train the mechanism that guesses the effects. The actions necessary to match guessed effects are, on the contrary, employed to train the decision mechanism about activities. In the last case, we have to use inverse rules, even though some problems arise when the inverse is indeterminate.

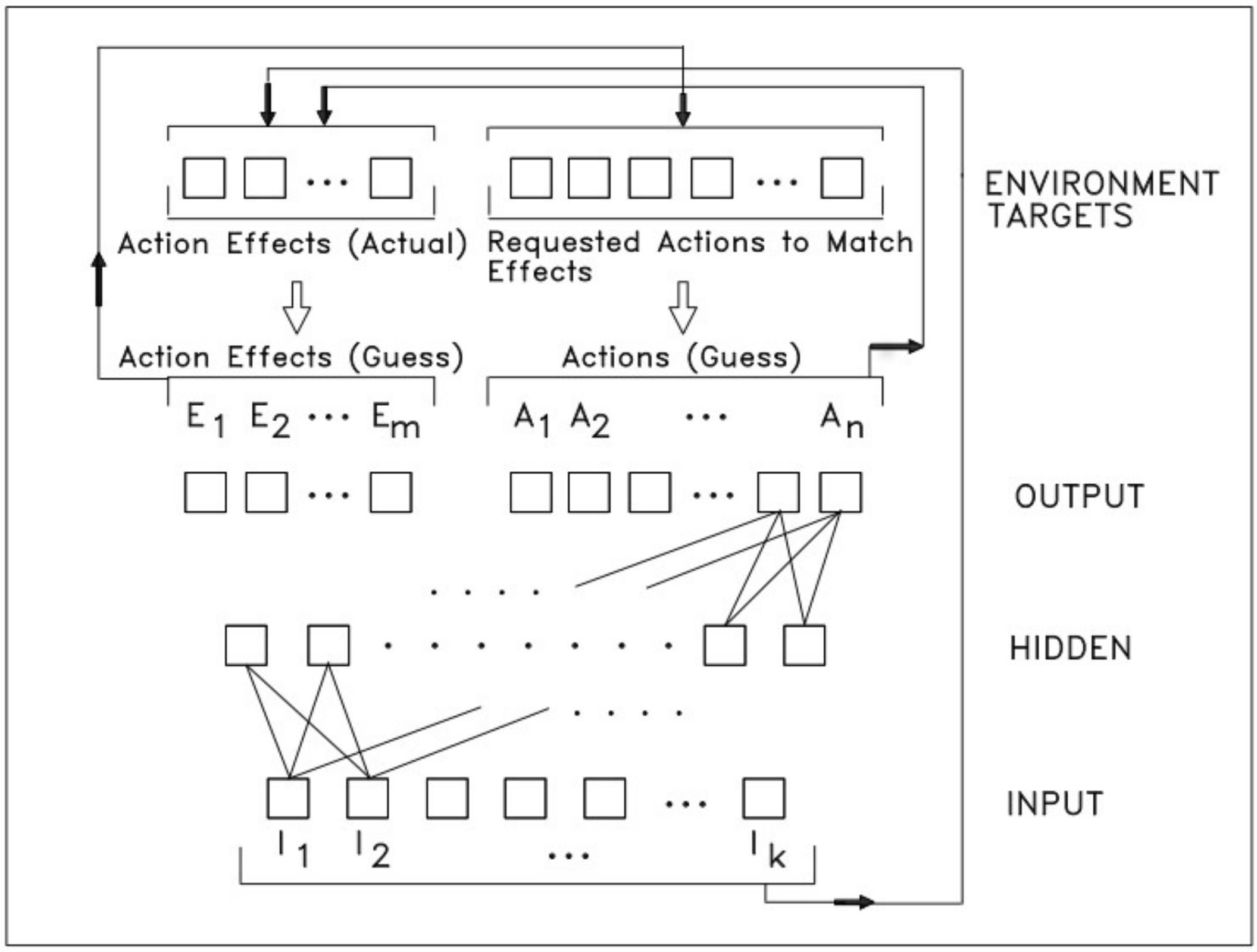

Figure 1 – The CT scheme

The CT method can appear to be related to Temporal Difference (TD) Learning of Barto and Sutton (Sutton 1988), using the differences between temporally successive predictions - or action outcomes - of the system, having a final target perfectly known at the end of the run. In the TD method we have a special and powerful case of true supervised learning, where an external teacher can suggest correct target values. TD, as CT, addresses the issue of consistent learning, but with delayed feedback founded upon a true target value; CT uses immediate tentative targets, self-generated and never corrected by an external teacher. The aim of CT is that of generating time paths for relevant variables, without any final or intermediate externally known objective, operating only with simple rules to adapt both behavior and predictions.

Observing Figure 1, we have: (i) the inputs of the model are mainly data coming from the environment or from other agents' behavior, (ii) they can be dependent or independent from the previous actions of the simulated artificial subject, (iii) targets are known only when actions take place.

Learning and acting take place in four steps each 'cycle'; a cycle is the sum of the four steps required to perform a full cycle of estimation of outputs and of backpropagation of errors, correcting the neural network weights (parameters).

Looking at Figure 1, the four steps can be introduced in the following sequence.

(1) Outputs of the artificial neural network: the actions to be accomplished, reported in the right side of Figure 1, and the effects of these actions, reported in the left side of the same figure, are guessed following inputs and network weights.

(2) Targets for the left side of the network: the targets for the effects supposed to arise from the actions, as guessed in the left side of the output layer in Figure 1, are figured out by the independently guessed actions. In this way, guesses about effects become closer to the true consequences of actual actions.

(3) Targets for the right side of the network: the differences measured in step (2) among targets and neural network outputs on the effect side can be inversely interpreted as starting points for action modifications, to match the guessed effects. So, they are used to build the targets for the mechanism that guesses the actions. Being the inverses of the formulas shown below often undefined, corrections are shared randomly among all the targets to be constructed; besides, when several corrections concern a target, only the one with the highest module is chosen. In this way, we would like to imitate the actual behavior of a subject requested of obeying to several independent and inconsistent commands: probably the most imperative, here the largest value, will be followed.

(4) Backpropagation: learning takes place, correcting weights in order to obtain guessed effects closer to the consequences of guessed actions, and guessed actions more consistent with guessed effects. Thus, we have two learning processes, both based upon the guesses of the elements of the opposite side of the network.

This double side process of adaptation, with interaction among agents and long-term learning (see above), ensures the emergences of non-trivial self-developed behavior outcomes.

We can now explain in a formal way the acting and learning structure of CTs, introducing a generic effect $E_1$ arising from two actions, named $A_1$ and $A_2$. The target for training a neural network able to reproduce this effect is:

$$E_1' = f(A_1, A_2) \qquad (1)$$

where $f$ is a definition, linking actions to effects on an accounting basis.

Our first goal here is to obtain an output $E_1$ (the guess made by the neural network) closer to $E_1'$, which is the correct measure of the effect of actions $A_1$ and $A_2$. The error related to $E_1$ is:

$$e = E_1' - E_1$$

or, by convention in neural network backpropagation, one half of the square of $E_1' - E_1$. To minimize the error, we backpropagate it through network parameters (weights).

The second (parallel) aim is that of ruling the actions, as outputs of our neural network, as more consistent with the outputs produced by the effect side. So, we have to correct $A_1$ and $A_2$ to made them closer to $A_1'$ and $A_2'$, which are actions consistent with the output $E_1$. We cannot figure out the targets for $A_1$ and $A_2$ separately. From (1) we have:

$A_1 = g_1(E_1', A_2)$ (2)

$A_2 = g_2(E_1', A_1)$ (3)

Choosing a value $\tau_1$ from a random uniform distribution whose support is the closed interval [0,1] and setting $\tau_2 = 1 - \tau_1$, from (2) and (3) we obtain:

$A_1' = g_1(E_1' - e \cdot \tau_1, A_2)$ (4)

$A_2' = g_2(E_1' - e \cdot \tau_2, A_1)$ (5)

Functions $g_1$ and $g_2$, being obtained from definitions that link actions to effects mainly on an accounting basis, usually have linear specifications; as a consequence, equations (4) and (5) generally give solutions that are globally consistent. The errors to be minimized are:

$a_1 = A_1' - A_1$

$a_2 = A_2' - A_2$

Eqs. 4 and 5 would not work as inversions of true dynamic functions, but they are used here as a simplified tool (thanks to the presence of random separation obtained by $\tau_1$ and $\tau_2$ values), to generate time paths for variables, without a priori or external suggestions.

When the actions determine multiple effects, they are included in multiple definitions of effects: those actions will be affected by several corrections; as reported in (3) above, only the largest absolute value is chosen.

Input and target variability, generated both in deterministic or random ways, is required to ensure the plausibility of the experiments, but is also necessary to ensure that are changing the outputs and the targets of the neural network under training. Lacking such a variability, on the basis of initial random weights of the network and following CTs, in most cases all outputs would be frozen at about 0.5, with perfect but merely apparent learning results.

**1.3 Short e long term**

With the proper variability, we repeat for a given number of cycles (e.g., real life days) the four steps introduced describing Figure 1. The learning activity follows the fourth step of each cycle and gives a sort of local adaptation to the changes of the environment.

Analyzing the changes of the weight (parameters of the neural network function) during that process, we can show that the matrix of weights linking input to hidden elements has little or no changes, while the matrix of weights from hidden to output layer changes in a relevant way. Only the hidden-output weight changes determine the continuous adaptations of the neural network responses to the environment modifications, while the output values of hidden layer elements stay almost constant.
This is the consequence of very small changes in targets (generated by CT method) and of a reduced number of learning cycles.
The resulting network is certainly undertrained. Consequently, the simulated economic agent develops a local ability to make decisions, but it has difficulties in coping with large environmental changes. This case resembles an actual consumer who is not able to determine her demand if prices change dramatically.
This is short term learning and it is opposed to long term learning, in analogy with the psychologists' distinction between short and long term memory: (a) The learning and acting phase produces neural agents continuously modifying their weights, in a local way, to adapt to environmental changes; (b) ex post, relearning the weights of the neural networks engaged in the experiment and using as data the historical records of the occurred events, we can obtain neural networks able to react, without subsequent learning, also to major changes in environmental conditions.
The second type of learning can also take place periodically upon a short segment of data (for example, every fifty cycles referring to the previous one hundred cycles, or upon a sample of the full historical data set, always with satisfying results).

## 2. Two examples

We introduce two examples[1] of short/long term learning and behaving activities: the first one is related to a simple experiment of motion, and the second one is related to an agent making two independent guesses, on the basis of a single input, the price $p_a$, which exogenously changes following a *sin* function plus a random noise. The two guesses are: the quantity $q_a$ of the good $C_a$ to be acquired; the expenditure $x$.

### 2.1 First example: movement

First example: the experiment on motion of agents foraging for food is built upon the following scheme.

[1] The reader can run the experiments with the code at https://github.com/terna/ct.

On a plane with ($x$, $y$) coordinates, the subject is initially in (*10*, *10*) while the food is fixed in (*0*, *0*). The neural network simulating the subject has as the inputs (which are the targets founded in the previous cycle):

$X(t\text{-}1)$, position in the x direction at the time $t$-1;

$Y(t\text{-}1)$, position in the $y$ direction at the time $t$-1;

$dX(t\text{-}1)$, step in the directions $x$, at time $t$-1 (bounded in the range ±1);

$dY(t\text{-}1)$, step in the directions $y$, at time $t$-1 (bounded in the range ±1).

Using the CT terminology, the same neural network produces as outputs two guesses about effects and two guesses about actions.

Positions $X(t)$ and $Y(t)$ have also the meaning of measuring the distance of the artificial subject from the food, set in (*0*, 0) (distance evaluated employing rectangular coordinates).

In Figure 2a we report the position of the agent in 200 cycles of acting and learning. The agent rushes toward the food but then stuck in the middle of the path.

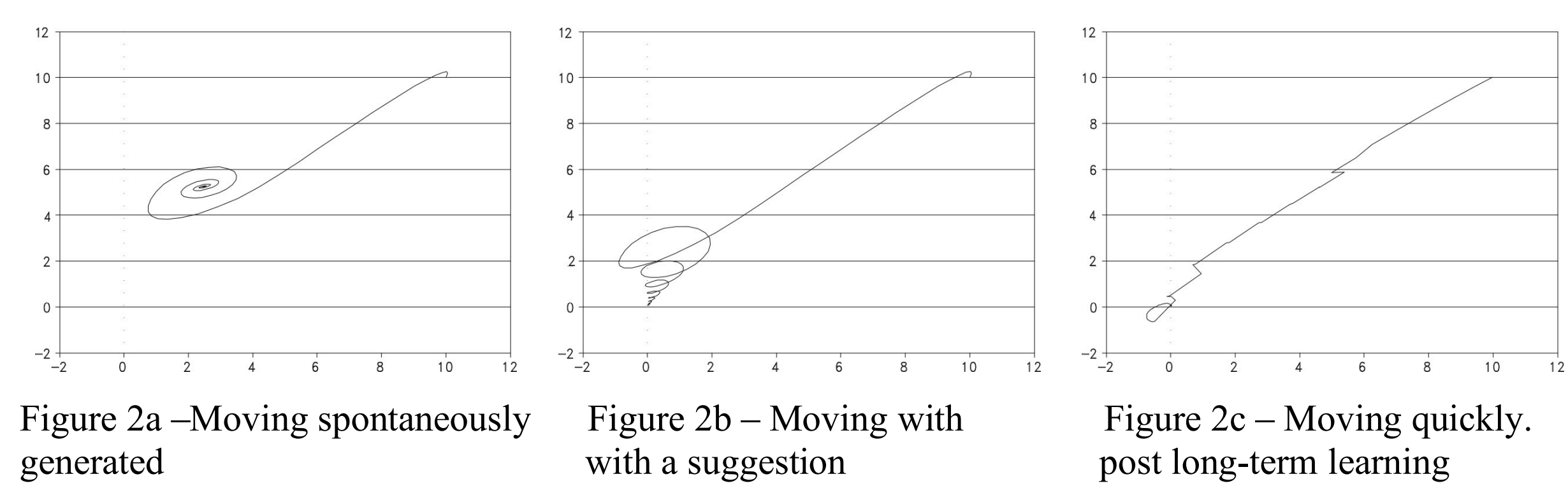

Figure 2a –Moving spontaneously generated

Figure 2b – Moving with with a suggestion

Figure 2c – Moving quickly. post long-term learning

The mechanism works in the following way: in the starting phase of the experiment, the neural network produces random outputs, but in a small interval around the central value between min and max. This effect is always present and is easily explained considering the consequence of the initial random choice of the weights, that gives on average a null sum of the inputs of the internal sigmoidal transformation. The initial guesses about the effects of the movement give estimated positions with large variability around the point (*0*, *0*) where the food is placed.

CTs immediately correct these wrong estimates, but they also correct the guesses about actions (the movements), to develop their consistency with the guesses of effects. In Figure 2a, the artificial agent moves in the right direction, but the process rapidly falls in a locking situation, with mutual consistency between effects and actions.

Imposing now an external objective on the side of the effects, that is the target of reducing in each cycle the distance from food to the 75% of the distance of the previous step, the food is easily gained, as reported in Figure 2b. We underline that no suggestion is introduced about the direction of the movement.

In Figure 2c we present again the case of Figure 2b, but with an artificial neural network whose weights come from a repeated learning process of 200 thousand cycles. The learning process is applied upon the full 200 cycles historical data of a single acting and learning run, in sequential order, with a momentum coefficient (usual value: 0.9).

The simulated agent reaches the food in few steps (less than one hundred), going directly toward it, despite some *realistic* (for the observer) uncertainty.

### 2.2 Second example: reacting to price changes

Second example: the simulated subject has to make two independent guesses, on the basis of a single input, the price $p_a$, which exogenously changes following a *sin* function plus a random noise. The guesses are: the quantity $q_a$ of the good *C*a to be acquired; the expenditure $x$. The cross-targets are easily established with:

$x' = p_a \cdot q_a$

$q_a' = q_a + (x - x')/p_a = x/p_a$

where, on one hand, $x'$ is the true expenditure, consistent with the decision $q_a$; on the other hand, $q_a'$ is the correct decision necessary to acquire the quantity of *A* consistent with the guess $x$. If $x' < x$, the guess of $q_a$ must be increased by the amount $(x - x')/p$ and vice versa.

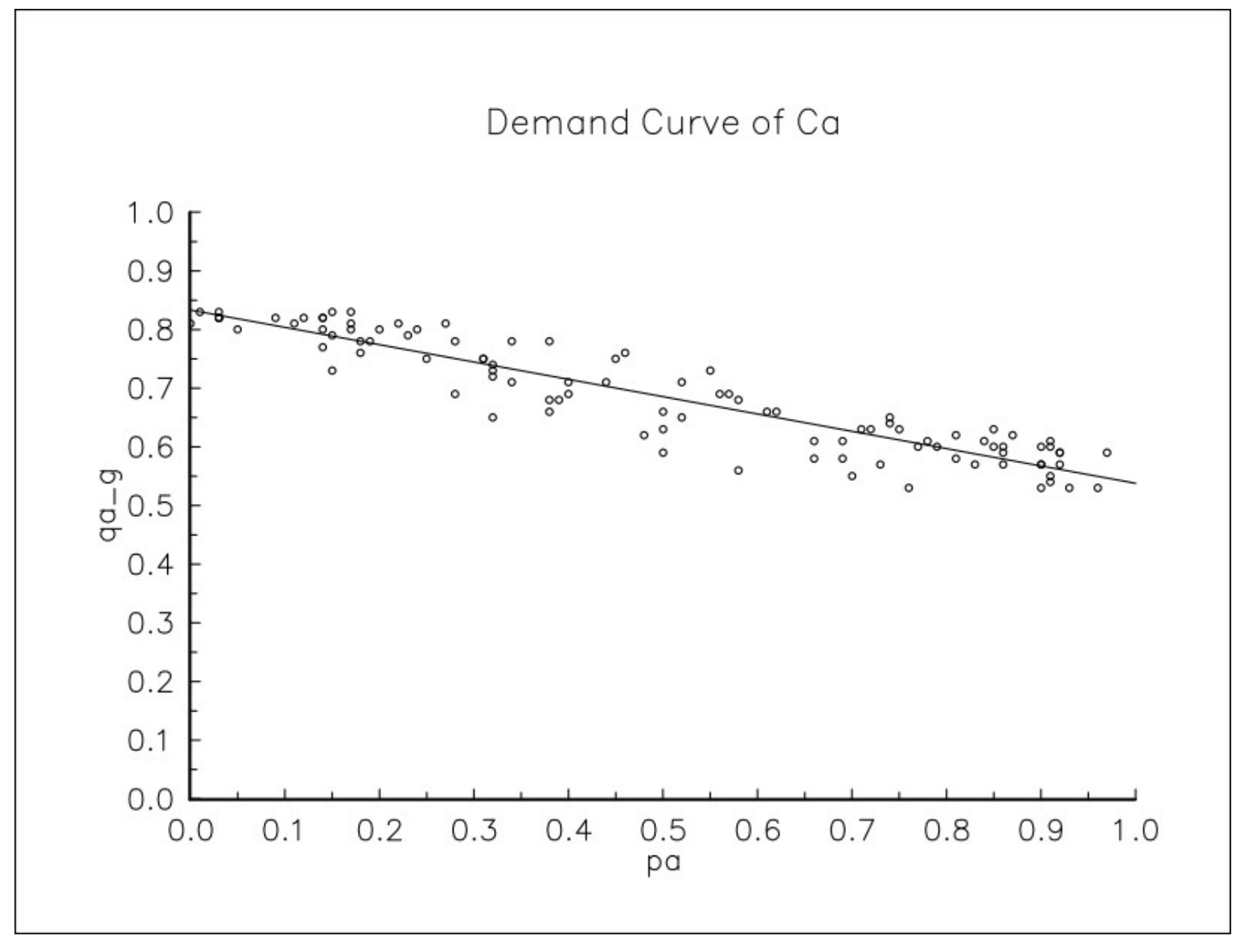

Figure 3 – A pseudo demand function, self-generated

In Figure 3 we have prices and guessed quantities (*qa_g*) for a thousand cycles of learning and acting; in the figure we have reported a sample of the data, based on a cycle every ten cycle. The simulated subject has self-developed the capability of reacting to price changes, smoothing the effects on its level of expense by quantity adjustment: what is emerging is a kind of demand curve without the need of an optimization hypothesis. Surprisingly, the mechanism, albeit very simple, is autonomously and endogenously developed by the model.

The dynamic: when the price changes, the cross-target mechanism determines two symmetrical corrections. Let us suppose that the price is increasing: the neural network outcomes, at the beginning of the learning process, are very conservative; consequently, in short term runs, the guesses about the expenditure (the effect) and about the quantity (the action) adapt themselves very slowly to the changing input. The expenditure evaluation will therefore be underestimated and corrected by increasing it; vice versa, the quantity, which has to be consistent with the guess about the expenditure, which is underestimated, has to be corrected by decreasing it. We have thus two symmetrical corrections.

Realistically, the results are not strictly exact nor deterministic. Is this a demand curve? In a strict sense it is not, but it describes very well a common behavior, perfectly consistent with the bounded rationality hypothesis.

## 3. Advanced techniques and perspectives

When we mention swarms of particles performing a distributed effort in order to perform a task, we cannot omit the humble random walkers (henceforth, RWs). Although very limited in their sophistication, and operational latitude, they can – and have been – successfully employed in network science to tackle a well-studied family of problems.

Contrary to standard PSO (Eberhart and Kennedy, 1995), where each "particle" represents a potential solution, searching the parameters space, we propose an algorithm aimed at converging to an equilibrium solution by means of a cooperative effort by multiple walkers, each one providing piecewise information adding up to a globally sensible response.

### 3.1 An Application to Complex Networks

In the "graph labeling problem", we observe a subset of labeled nodes on a network structure of the form $G(U, E_w)$, where U is the set of nodes and $E_w$ is the set of weighted, possibly directed edges. The purpose of the process is to appropriately label the entirety of nodes, based on some classification paradigm. Subramanya and Talukdar, 2014 makes for an excellent introductory read.

As a prime example, and given the prompt availability of such data, we may observe an online social network of some kind, where neighboring individuals/nodes – and nodes reachable in a few steps in general – will tend to exhibit homophily: similar individuals will be more likely to be linked. Intuitively put, it can be argued that "diffusing" the available pieces of information (labels) around, within a given topological distance from a labeled node, may be informative of the characteristics of the anonymous nodes too. For the sake of simplicity, let's also assume a binary categorical labeling, where a few nodes are known to belong (are labeled) to either the "red" or the "blue" team. Roughly speaking, inferring the team of the anonymous nodes can be performed by sending RWs along the network, counting the number of visits payed to each node by class of the RW's originating node, and observing which kind of visiting agent is prevalent.

### 3.2 Nodes labelling and graph-based semi-supervised learning (G-SSL)

Multiple algorithms in this vein have been proposed (see, as a prominent example, PageRank, in Brin and Page, 1998, or de Nigris *et al.*, 2017 for a comprehensive review of applications), which can be solved either analytically, with a closed form solution (the analytical approach, though, comes with a major scaling problem, due to the computational cost of inverting a very large matrix), or as a – case in point – random walk simulation. We take the latter approach, in order to present and solve the problem as a distributed, independent effort by multiple, rationally bounded entities gathering and exploiting information locally (this approach also preserves scalability up to very large-scale datasets).

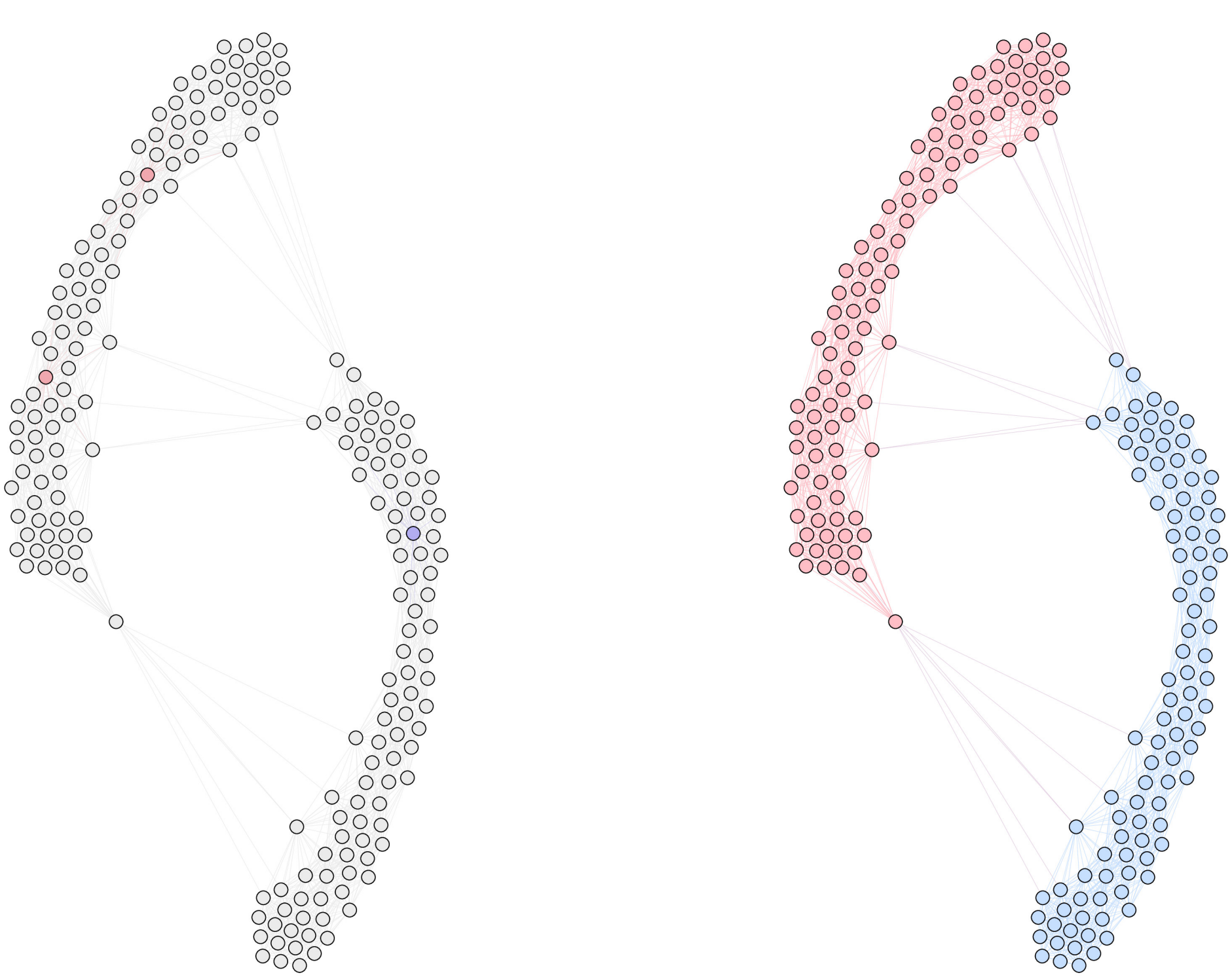

Figure 4 - Seed nodes on a partially labeled graph

Figure5 - Node classification by CT-GSSL

### 3.3 Random Walkers working as a learning system

A set of RWs is created and sent around the network, starting from labeled "seed" nodes (see Figure 4, seed nodes colored in red and blue, respectively), and hopping between nodes, in turn increasing a class-specific visits counter for everyone: every time a node is visited by a RW, it increases its likelihood to belong to the former's class accordingly. Visited nodes will be considered to belong to the class of the prevailing seed color, and consequentially classified. A successful classification would appear as the one presented in Figure 5.

In order to attribute due relevance to the class of the RW's originating node, the concept of "restarting probability" is introduced: with a given probability $1-p_r$, RWs can either keep on with their exploration, or (with probability $p_r$) go back to their originating node, ready to travel along a brand-new itinerary taming the diffusive process and keeping it from converging to the stationary distribution of a RW.

The shortcomings of existing algorithms are known and include the misclassification of nodes originating from the diffusion of walkers through spurious links connecting the classes and is apparent when benchmarking them on graphs with known, and well-defined, planted partitions (again, refer to Subramanya and Talukdar, *op. cit.*). Multiple efforts to improve the algorithms accuracy have been undertaken by the research community, and a vast literature exists on the topic.

### 3.4 A learning classification system

The following exercise is meant to constitute an attempt at going beyond the existing, zero-intelligence, diffusive techniques, allowing RWs to acquire an exploration strategy which is not constrained by any a priori assumption on the characteristics of the problem. Let us envision a sophisticated kind of RW which is endowed with some sort of directed intelligence, biasing the process based on some learnt heuristics instead of on some external, arbitrarily chosen, metric.

At each iteration, each RW is required to move onto a node picked out of its first neighbors; the intended aim is to keep moving onto nodes belonging to the same partition, given the information available, and refraining from crossing "partitions boundaries". The CT is fed (input neurons) with data on both the originating ($u$), the potential destination ($v$) node, and eventually (in the case of a weighted graph) the link weight ($w_{uv}$); on the output side a positive response (+1) will be elicited by a successful transition (on a node belonging to the same partition), whereas a *faux pas* towards a different partition will result in a null value (0).

Inputted data consist of local information – a vector of characteristics (metadata) of nodes *u* and *v* – and network-specific metrics, which may (e.g. *Betweenness Centrality*) or may not (e.g. *Degree*) incorporate structural-topological, global aspects.
For each RW, the target (*à la* CT) is supposed, as mentioned earlier, not to trespass communities boundaries, that is, to keep exploring homogeneous areas of the network; from the ANN perspective, being awarded a positive value on the output node.
In the training phase, a benchmark network is used to expose the cross-target-based, neural-network-powered agents to the "environment" to which they are supposed to adapt, going through the 4 steps illustrated in the paragraph "The Learning Phases", above. Once consistency and an appropriate strategy have been developed, eventually checking for individual idiosyncrasies, experiments can be performed on partially labeled networks, assessing the accuracy of the process. Finally, primed RWs, validated for the task, can be put to work on different networks.
A particularly intriguing application involves evolving temporal networks (such as contact networks), where new nodes appear: specifically, trained agents may be employed to incrementally label them as the network evolves.
As a final consideration, we dispose of a scalable, intrinsically parallel algorithm, capable of achieving high-accuracy classification of nodes and, as a bonus, making the best use of available information in order to inform the exploration process.